\documentclass[aps,prb,superscriptaddress,twocolumn]{revtex4-1}
\usepackage{bm}
\usepackage{hyperref}
\usepackage{graphicx}
\renewcommand{\Im}{\mathop{\text{Im}}}
\begin{document}

\title{Boundary Conditions and Surface States Spectra in Topological Insulators}

\author{V.V. Enaldiev}
\affiliation{Kotelnikov Institute of Radio-engineering and Electronics of the Russian Academy of Sciences, 11-7 Mokhovaya, Moscow, 125009 Russia}

\author{I.V. Zagorodnev}
\affiliation{Kotelnikov Institute of Radio-engineering and Electronics of the Russian Academy of Sciences, 11-7 Mokhovaya, Moscow, 125009 Russia}

\author{V.A. Volkov}
\affiliation{Kotelnikov Institute of Radio-engineering and Electronics of the Russian Academy of Sciences, 11-7 Mokhovaya, Moscow, 125009 Russia}
\affiliation{Moscow Institute of Physics and Technology, Institutskii per. 9, Dolgoprudny, Moscow Region, 141700 Russia}

\date{\today}

\begin{abstract}

We study spectra of surface states in 2D topological insulators (TIs) based on HgTe/(Hg,Cd)Te quantum wells and 3D Bi$_2$Se$_3$-type compounds by constructing a class of feasible time-reversal invariant boundary conditions (BCs) for an effective ${\bf k}{\bf p}$-Hamiltonian and a tight-binding model of the topological insulators. The BCs contain some phenomenological parameters which implicitly depend on both bulk Hamiltonian parameters and crystal potential behavior near the crystal surface. Space symmetry reduces the number of the boundary parameters to four real parameters in the 2D case and three in the 3D case. We found that the boundary parameters may strongly affect not only an energy spectrum but even the very existence of these states inside the bulk gap near the Brillouin zone center. Nevertheless, we reveal in frames of the tight-binding model that when surface states do not exist in the bulk gap in the Brillouin zone center they cross the gap in other points of the Brillouin zone in agreement with the bulk-boundary correspondence.

\end{abstract}

\maketitle

\section{Introduction}

A topological insulator (TI) is an intriguing quantum state of matter which is insulating in its interior and possesses surface states (SSs) crossing the bulk energy gap \cite{bib:Hasan,bib:Qi,bib:Franz}. The existence of the SSs is provided by a $Z_2$ topological invariant and the bulk-boundary correspondence. The $Z_2$ invariant is a characteristic of the bulk band structure of a crystal and it comprises no information about the crystal surface. The bulk-boundary correspondence asserts that the odd numbers of pairs of gapless SSs should arise \cite{bib:Hasan,bib:Qi,bib:Franz,bib:Isaev,bib:Graf} at the interface between two crystals with different values of the $Z_2$ invariant, for instance, between vacuum (with the $Z_2$ index equals zero) and the TI (with the $Z_2$ index equals one), but it does not characterize the dispersion of the SS spectra. 
 
Topological SSs have been studied in a number of papers in tight-binding approximations \cite{bib:Isaev,bib:Persh} and in envelope function approximation \cite{bib:Qi,bib:Linder,bib:Zhou,bib:Liu, bib:Medhi, bib:Menshov,bib:Tugushev}. Nevertheless in any approach spectra of the SSs should depend on details of crystal-vacuum interface structure. To take it into account one usually chooses  appropriate boundary conditions (BCs) for a wave function of topological SSs in a specific model. Most of authors use open BCs (wave function vanishes at the surface) which guarantee massless Dirac spectrum of SSs in TIs. However, open BCs are not highlighted by nature and the other BCs might be realized. There are few papers that address the problem of BCs for a wave function of SSs at the surface. 

Within a 3D tight-binding model of TI \cite{bib:Isaev} it was shown that a strong topological insulator cannot be transformed into a trivial insulator by means of varying BCs. But that tight-binding model is a toy model and does not describe any really existing 3D TIs. Therefore, it is important to study BC influence on SSs in frames of more realistic \textbf{kp}-Hamiltonian model though it is only valid in a vicinity of Brillouin zone (BZ) center. Refs.\cite{bib:Medhi, bib:Menshov,bib:Tugushev} analyzed dispersion dependence of SSs on the BCs for envelope functions (eigenfunctions of the \textbf{kp}-Hamiltonian). In this approach one should exploit a matrix form of BCs with some unknown parameters \cite{bib:Ivchenko} that connect envelope functions and their derivatives on the interface. The boundary parameters are determined by the interface structure (on the atomic scale) as well as bulk band parameters of materials.  As microscopic details of a real interface are unknown it is involved problem to calculate values of the boundary parameters. More reliable method to determine the parameters is to extract them from experiment (for instance, in Ref.\cite{bib:Devizorova} it was done for the most studied GaAs/AlGaAs interface).  

Aim of our paper is to demonstrate that the boundary parameters play crucial role not only for the spectrum of topological SSs but also the very existence of SSs in bulk gap near Brillouin zone center, where envelope function approximation is valid. In addition, we show that this is in agreement with the bulk-boundary correspondence. To that end we consider a 2D tight-binding model of TI with general BCs at the edge. Tuning the boundary parameters in the model we can modify spectra of SSs over the full BZ, for instance, so that they cross the bulk gap at the edge of the BZ.

The paper is organized as follows. First, we derive general BCs for the envelope functions in HgTe/(Hg,Cd)Te quantum wells and in Bi$_2$Se$_3$-type 3D TI and analyze edge state (ES) spectra in 2D TI and SS spectra in 3D TI. Then we study an effect of general BCs within a tight-binding model of 2D TI that allows us to clarify behavior of the SS spectra over the full Brillouin zone (BZ).    


\section{2D topological insulators}\label{Section_2D_TI}

The phase of 2D TI was obtained in HgTe/(Hg,Cd)Te quantum wells with a certain thickness and composition \cite{bib:Bern}. The electronic spectrum in the quantum wells near the critical thickness is described by the effective 2D Hamiltonian \cite{bib:Qi,bib:Raichev}
\begin{equation} \label{2DHam}
	H_{2D} = \sigma_0 \otimes \left( m(k)\tau_z -dk^2 \tau_0 + vk_y\tau_y \right) + vk_x \sigma_z \otimes \tau_x.
\end{equation}
Here, $\hbar{\bf k} = \hbar(k_x, k_y)$ is the 2D momentum, $k^2 = k_x^2+k_y^2$, $v/\hbar$ is the effective speed of light ($v>0$), $\sigma_{0,x,y,z}$ and $\tau_{0,x,y,z}$ are the Pauli matrices in the standard representation acting in spin and orbital subspaces, and $\otimes$ is the symbol for the direct product. The parameters $b,d<0$ are responsible for the dispersion of the mass $m(k)=m_0-bk^2$, leading to the modification of the Dirac spectrum $E = \pm \sqrt{m_0^2+v^2k^2}$, and are expected to have significant importance for the appearance of the TI ($m_0<0$ in TI phase), Ref. [\onlinecite{bib:Qi}].

To use the Hamiltonian (\ref{2DHam}) in a restricted area, it should be supplemented by the BCs at the edge of the system. To do this, we use general physical requirements that significantly restrict the form of the BCs. First of all, since the Hamiltonian (\ref{2DHam}) is of the second order in the momenta, we assume that the BCs are a linear combination of the wave function and its first derivative
\begin{equation} \label{2D_genBC}
	\left.\left( F\partial_{\bf n}\psi + G\psi \right)\right|_S = 0,
\end{equation}
where ${\bf n} = (\cos\alpha, \sin\alpha)$ is the outer normal to the edge, and $G$ is an arbitrary $4 \times 4$ matrix. For convenience, the matrix $F$ can be selected as follows:
\begin{equation} \label{2D_generalBC}
	F = \frac{b}{v}\sigma_0\otimes\tau_0 + \frac{d}{v}\sigma_0\otimes\tau_z.
\end{equation}
Second, we use the Hermiticity of the Hamiltonian in a restricted area. Therefore, we perform partial integration of $<\psi| H_{2D} |\varphi>$ for arbitrary wave functions $\varphi$, $\psi$, and equating the surface term to zero, we find the restriction
\begin{equation}\label{HermitianG}
	G^+\sigma_0\otimes\tau_z - \sigma_0\otimes\tau_zG - i\sigma_z\otimes\bm{\tau} \bm{n}   = 0.
\end{equation}
This indicates that Eq. (\ref{HermitianG}) implies the absence of current normal to the edge. Next, we take into consideration the time-reversal symmetry with respect to the operator
\begin{equation}
	\hat{T} = i\sigma_y\otimes\tau_0 \hat{K},
\end{equation}
where $\hat{K}$ is the complex conjugate. Applying commutation of the operator $\hat{T}$ with $G$ and Eq. (\ref{HermitianG}) we obtain the matrix $G$ of the most general form:
\begin{equation} \label{2D_BCmatrix}
	\left( \begin{array}{cccc} 
		g_1 & g_2+ig_3 & 0 & g_5+ig_6\\
		i(e^{i\alpha}+g_3)-g_2 & g_4 & g_5+ig_6 & 0 \\
		0 & ig_6-g_5 & g_1 & g_2-ig_3 \\
		ig_6-g_5  & 0 & -i(e^{-i\alpha}+g_3)-g_2 & g_4
	\end{array} \right),
\end{equation}
where $g_{\overline{1,6}}$ are dimensionless real phenomenological boundary parameters, which depend both on the behavior of the crystal potential near the edge and the bulk Hamiltonian parameters. Their values should be determined by microscopic calculations or by experiments. The open BCs correspond to $g_1,g_4\rightarrow\infty$. The ''natural'' BCs \cite{bib:Medhi} correspond to $g_2 = i\cos\alpha+(\sin\alpha)/2$, $g_3 = -(\cos\alpha)/2$, and the other parameters equal zero.

Consider two important particular cases. The first case occurs when boundary potential does not mix the spin of electrons. Since the upper (lower) two components of the wave function correspond to spin-up (-down), this case results in $g_5 = g_6 = 0$. The second case is when the edge, assumed as $x=0$, possesses spatial inversion, e.g. $y\rightarrow-y$, which is described by the operator
\begin{equation}
	I_y = \sigma_x\otimes\tau_z\hat{i}_y,
\end{equation}
where $\hat{i}_y$ is the coordinate inversion $y\rightarrow - y$. The commutation of the operator $I_y$ with $G$ gives $g_2=g_5=0$ in (\ref{2D_BCmatrix}) with $\alpha = 0$.

\begin{figure}
	\includegraphics[width=8cm]{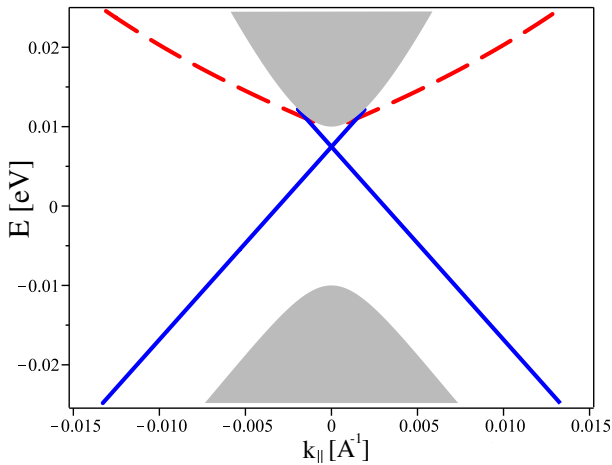}
	\caption{\label{Fig:SSspectr2DTI}The electron spectrum $E(k_{||})$ of semi-infinite HgTe/(Hg,Cd)Te quantum well. The shaded region corresponds to the continuous spectrum (bulk solutions). The edge states are described by solid curves for open boundary conditions ($g_1 = g_4 = \infty$, $g_2 = g_3 =0$), and dashed curves for $g_1 = -2$, $g_4 = \infty$ (all the other boundary parameters equal zero for both curves). Thus, the edge states vanish from the gap of the TI for some values of the boundary parameters. Nevertheless there are edge states in the bulk gap but they cross the bulk gap at large momenta that are beyond the scope of the envelope function approximation, see Fig.\ref{Fig:SSspectrLattice2DTI}.}
\end{figure}

To analyze the ES spectra, we solve the Schroedinger equation $H_{2D}\psi=E\psi$ on the half-plane $x>0$ with the BCs (\ref{2D_genBC}) and (\ref{2D_BCmatrix}). The wave vector along the edge, $k_{||}$, is a good quantum number. The bulk solutions are located in the energy region $|E-dk_{||}^2| \geq \sqrt{(m_0-bk_{||}^2)^2+(vk_{||})^2}$, while the ES solutions are situated outside the region. The wave function of the ESs contains two exponents, $\exp(-\kappa_1x)$ and $\exp(-\kappa_2x)$. The energy spectrum of the ESs is shown in Fig.\ref{Fig:SSspectr2DTI}. The spectrum of the ESs corresponding to the open BCs is shown by solid curves. In this case, the ESs have a strictly linear dispersion and their decay lengths in the bulk gap are estimated as $1/\kappa_1 \approx \sqrt{b^2-d^2}/v \approx 1$ nm (which is comparable to the atomic spacing) and $1/\kappa_2 \approx 50$ nm with slight dependence on $k_{||}$ (the bulk parameters was taken from Ref.[\onlinecite{bib:Qi}] for $m_0 = -0.01$ eV). However, if the parameter $g_1$ significantly decreases the ESs pushed out of the band gap, see the dashed curves. To better understand this extraordinary behavior of the ESs, we may note that usually the parameters $b$ and $d$, describing the quadratic in $k$ terms of the Hamiltonian (\ref{2DHam}), are small enough we therefore consider a limit $b,d\rightarrow 0$.

\begin{figure} 
	\includegraphics[width=8cm]{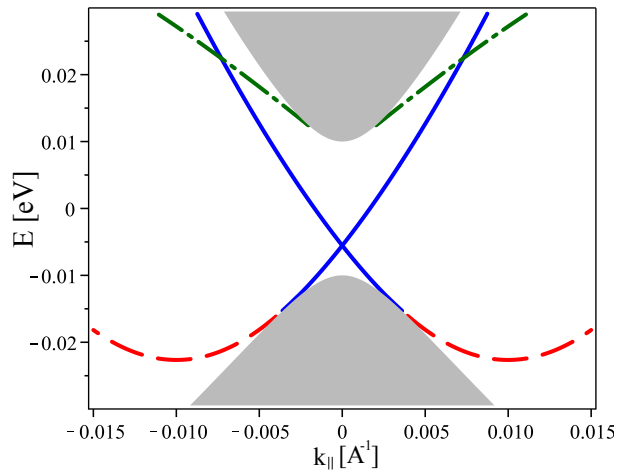}
	\caption{\label{Fig:SSspectr2DTIDirac}The electron spectrum of semi-infinite HgTe/(Hg,Cd)Te quantum well, where the boundary is described by the single parameter (the BCs (\ref{2D_genBC}),(\ref{2D_BCmatrix}), and (\ref{2DTI-DiracBC})). The shaded region corresponds to the continuous spectrum. The edge states are described by solid curves for $a_0 = -2$, dashed curves for $a_0 = 0.3$ and dash-dotted curves for $a_0=2$.}
\end{figure}

At the limit, the bulk spectrum has exactly the 2D Dirac form and it is instructive to compare our results with known ES spectra for the 3D Dirac equation \cite{bib:VolkPin} with $k_z=0$. We should point out that in this case, $\kappa_1 \rightarrow \infty$ for open BCs, so that one of the ES decay length is small by comparison to the atomic distance. The part of the wave function $\exp(-\kappa_1x)$ can not be described by the effective Hamiltonian (\ref{2DHam}), but it might be included in the BCs. We use the unitary transformation
\begin{equation}
	U = \frac{1}{2}\left( 
	\begin{array}{cc}
		\sigma_0-\sigma_z & -\sigma_x-i\sigma_y\\
		\sigma_0+\sigma_z & \sigma_x-i\sigma_y
	\end{array}
	 \right),
	 \quad \widetilde{\psi} = U\psi
\end{equation}
of the Hamiltonian (\ref{2DHam}) and reduce it to another form:
\begin{equation}
	U H_{2D} U^{-1} = m_0\sigma_z\otimes\tau_0+v\sigma_x\otimes\bm{\tau}{\bf k}.
\end{equation}
This is merely one of the standard forms of the 3D Dirac equation with $k_z = 0$. The time-reversal invariant BCs for the Dirac Hamiltonian derived from Hermiticity have the following form \cite{bib:VolkPin}:
\begin{equation} \label{DiracBC}
	\left. \left( -ia_0 \frac{\sigma_0+\sigma_z }{2} \otimes \bm{\tau}{\bf n} + \frac{\sigma_x+i\sigma_y}{2}\otimes\tau_0 \right) \widetilde{\psi}\right|_S = 0,
\end{equation}
where $a_0\in(-\infty,\infty)$ is a real dimensionless phenomenological parameter. Comparing the transformed BCs (\ref{2D_genBC}), and (\ref{2D_BCmatrix}) for the 2D TI when $b,d=0$ (in this case the normal derivative vanishes) with Eq. (\ref{DiracBC}), we obtain
\begin{equation}\label{2DTI-DiracBC}
	ia_0e^{i\alpha}g_1=(g_2+ig_3),\, -\frac{g_4}{a_0} = 1 + e^{-i\alpha}(g_3+ig_2),\, g_5=g_6=0.
\end{equation}
The remaining uncertainty in the parameters $g_i$ for fixed $a_0$ does not influence the ES spectra. The electronic spectrum of the Hamiltonian (\ref{2DHam}) with the BCs on the half-plane $x>0$ is shown in Fig.\ref{Fig:SSspectr2DTIDirac}. For small $k$, one may neglect the quadratic terms in $k$ in the Hamiltonian (\ref{2DHam}). Therefore the ES spectrum of the Dirac equation \cite{bib:VolkPin,bib:VolkId}
\begin{equation}\label{DiracSSspectra}
	E = \frac{1-a_0^2}{1+a_0^2}m_0 \pm \frac{2a_0}{1+a_0^2}vk_{||},\quad \frac{2a_0}{1+a_0^2}m_0 \mp \frac{1-a_0^2}{1+a_0^2}vk_{||} < 0
\end{equation}
is equivalent to that of the 2D TI. For large $k$, the quadratic term dominates, but the Hamiltonian (\ref{2DHam}) and the BCs have limited application in the area of $k$. The key feature of the obtained ES spectrum is that if $m_0 a_0<0$ then the ESs are in the bulk gap near $\bm k = 0$, while if $m_0 a_0>0$ there are no ESs near $\bm k = 0$. This means that massless ESs (Dirac cone) can be found in the gap of the TI phase ($m_0<0$) if $a_0>0$, as well as in the trivial phase ($m_0>0$) if $a_0<0$. However, in the former case the ESs are not topologically protected.

To finally persuade the reader, let us simplify a general dispersion equation that describe the ES spectra of the Hamiltonian (\ref{2DHam}) on the half-plane with the BCs (\ref{2D_genBC}), and (\ref{2D_BCmatrix}), in a case where $d\approx 0$ and $b$ is small. Without loss of generality, we can consider the BCs which do not mix the spin of electrons (i.e., $g_2=g_5=0$) and possess spatial inversion $y\rightarrow -y$ (i.e., $g_6=0$). Then, in the first order in the parameter $b/v$, one can obtain the spectra of ESs again in the form (\ref{DiracSSspectra}), but with
\begin{equation}
	a_0 = - \frac{g_1(g_4+1)+(1+g_3)^2}{g_3^2+g_4(1+g_1)}.
\end{equation}

Hence, once again, the ES spectra depend significantly on boundary parameters and the inversion of mass is not a necessary or sufficient condition for the existence of the ESs in the bulk gap at small momenta when the $kp$-approximation is valid.

Thus, we have demonstrated that the open BCs for the Hamiltonian (\ref{2DHam}) are not the only ones possible. The solutions of the effective Hamiltonian (\ref{2DHam}) with the open BCs have a limited range of applicability due to the inadequate profile of the wave function for small $b$ and $d$. In this case, it is more suitable and much easier to use the Dirac equation with the BCs (\ref{DiracBC}), which contain only one boundary parameter, $a_0$. We stress that boundary parameters may remove ESs from the gap at small momenta. Further (sec.\ref{lattice_model}) we will show that in this case the SSs will cross the bulk gap near edges of the BZ in agreement with the bulk-boundary correspondence.


\section{3D topological insulators}\label{Section_3D_TI}

Here, we consider a simplified and isotropic \textbf{kp}-Hamiltonian of 3D TI Bi$_2$Se$_3$ near the center of the BZ. This is the 3D Dirac Hamiltonian with the momentum depending mass term\cite{bib:Liu} 
\begin{equation}\label{3D_Ham}
H_{3D}=m({\bf k })\sigma_0\otimes\tau_z + v(\bm{\sigma} \cdot {\bf k}) \otimes\tau_x
\end{equation}  
where $m({\bf k})=m_0+b{\bf k}^2$, ${\bf k}=(k_x,k_y,k_z)$ is the three-dimensional momentum vector, $v,m_0$ and $b$ are the model parameters. Let us consider BCs for eigenfunctions of the Hamiltonian (\ref{3D_Ham}). As was noted in the previous section, the most general BCs have the following form
\begin{equation}\label{BC_general}
	\left.\left( \frac{b}{v}\partial_{\bm{n}}\Psi-Q\Psi \right) \right|_S=0
\end{equation}
where $Q$ is a $4\times 4$ matrix with complex parameters that phenomenologically describe the microscopic surface potential, and $\Psi=(\Psi_1,\Psi_2,\Psi_3,\Psi_4)^{T}$ is a bispinor consisting of the four envelope functions. 

One can reduce the number of parameters in the BCs by applying symmetry considerations. Below, we consider a surface with outer normal ${\bf n}=(0,0,-1)$, and assume that the BCs (\ref{BC_general}) possess symmetries of a semi-infinite Bi$_2$Se$_3$ crystal with the (111) surface. They include the rotation $R_{3,z}$ of the coordinate system on the angle $2\pi/3$ about the $z$ axis, time-reversal symmetry $T$ and reflection $I_{x}$ in $yz$ plane. For the Hamiltonian\cite{bib:Liu} (\ref{3D_Ham}), these operations of the symmetry are represented by $4\times 4$ matrices $\hat{R}_{3,z}=e^{i\sigma_z\otimes\tau_0\pi/3}$, $\hat{T}=i\sigma_y\otimes\tau_0\hat{K}$, and $\hat{I}_{x}=i\sigma_x\otimes\tau_z$. The invariance of the BCs upon the symmetry operation, e.g., $I_{x}$, means that relation (\ref{BC_general}) with the same
boundary condition matrix $Q$ can be applied to the transformed bispinor
\begin{equation}\label{symmetry_operation} 
\begin{array}{c}
I_{x}\hat{\Gamma}(\bm{r})\Psi(\bm{r})|_{ S}=\hat{I}_{x}\hat{\Gamma}(I_{x}^{-1}\tilde{\bm{r}})\hat{I}_{x}^{-1}\hat{I}_{x}\Psi(I_{x}^{-1}\tilde{\bm{r}})|_{S}= \\
\hat{\Gamma}(\tilde{\bm{r}})\tilde{\Psi}(\tilde{\bm{r}})|_{S}=0,
\end{array}
\end{equation}
where the boundary operator is denoted \mbox{$\hat{\Gamma}(\bm{r})=b\partial_{\bm{n}}/v-Q$}. Applying the symmetry operations together with the requirement of the Hermiticity of the Hamiltonian (\ref{3D_Ham}) in a half-space $z\geq 0$, leads to the following form for $Q$: 
\begin{equation}
Q = \left(
\begin{array}{cccc}
q_1 & iq_2+\frac{i}{2}& 0 & 0 \\
iq_2-\frac{i}{2} & q_3 & 0 & 0 \\
0 & 0 & q_1 & -iq_2-\frac{i}{2}\\
0 & 0 & -iq_2+\frac{i}{2} & q_3
\end{array}
\right)
\end{equation}
where $q_1, q_2$ and $q_3$ are real phenomenological boundary parameters which implicitly characterize the microscopic surface structure. The matrix Q would be exactly the same if we used infinite order rotational symmetry around $z$-axis instead of $\hat{R}_{3,z}$. BCs similar to the BCs (\ref{BC_general}) were also considered in Ref.[\onlinecite{bib:Menshov}], but that matrix $Q$ contains 16 arbitrary complex parameters. By means of the symmetry analysis we decreased the number of the parameters to the three real parameters. The authors of the Ref.[\onlinecite{bib:Menshov}] analyzed the case of arbitrary $q_1$ and $q_3$ with $q_2=0$ and found that the Dirac point merged in the bulk spectrum at certain values of $q_1$ and $q_3$. 

\begin{figure}
\includegraphics[width=8cm]{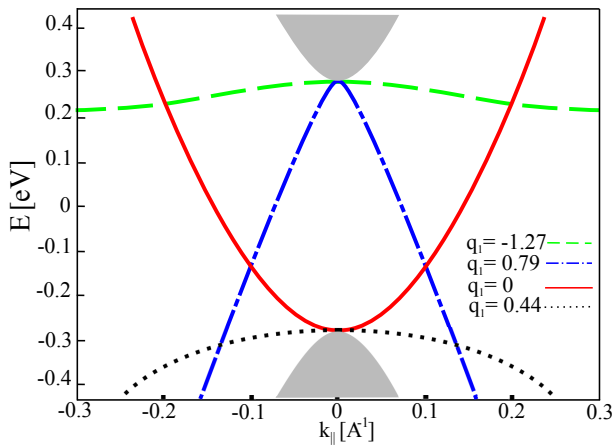}
\caption{\label{Fig:3Dspectrum} Effect of boundary parameters on the SS energy spectrum of the Bi$_2$Se$_3$-type 3D TI. Each curve corresponds to a definite value of $q_1$. Shaded regions are the bulk states. Variation of the boundary parameters can lead to the removal of the SSs or their Dirac point from the bulk band gap. The spectrum was calculated at $q_2=1/2$, $q_3=-q_1$, $m_0=-0.28$ eV, $b=6.86$ eV$\cdot$\r{A}$^2$, $v=5$ eV$\cdot$\r{A}.}
\end{figure}

We first study the case $q_1=-q_3$ with arbitrary $q_2$, as it captures the main results of this section. We can find the bispinor $\Psi$ satisfying the Schroedinger equation $H_{3D}\Psi=E\Psi$ in the following form
\begin{equation}\label{3D_wave_function}
\begin{array}{l}
\Psi=h_1(k_{||},E)e^{-\kappa_1z}+h_2(k_{||},E)e^{-\kappa_1z}+ h_3(k_{||},E)e^{-\kappa_2z} \\
\qquad\qquad+h_4(k_{||},E)e^{-\kappa_2z} \\
\end{array}
\end{equation}
\begin{equation}
\begin{array}{l}
\kappa_{1,2}(k_{||},E)=\frac{1}{\sqrt{2}b}( v^2+2b^2k_{||}^2+2m_0b\mp \\
\sqrt{v^4+4v^2m_0b+4E^2b^2} )^{1/2}, 
\end{array}
\end{equation}
where $k_{||}=|\bm{k}_{||}|$. Following Ref.[\onlinecite{bib:Liu}] we imply $m_0<0$, $b>0$ for the TI phase. The BCs (\ref{BC_general}) determine the four-component vectors $h_{1,2,3,4}(k_{||},E)$ and the SS spectrum. Through algebra, one derives a dispersion equation, which for $k_{||}=0$ reduces to
\begin{equation}\label{Dirac_point_eq}
v^2\kappa_1\kappa_2\left[ \left (q_2+\frac{1}{2}\right )F_- +\left (q_2-\frac{1}{2}\right )F_+ \right] - F_+F_-=0,
\end{equation}
\begin{equation}
\begin{array}{l}
F_{\pm}=\frac{v^2}{b}\left (q_2\mp\frac{1}{2}\right )(m_0\pm E)\mp \\ \left[ \frac{v^2}{b}\left ( q_2^2-q_1^2-\frac{1}{4}+\frac{b^2}{v^2}\kappa_1\kappa_2 \right )(m_0\pm E+b\kappa_1\kappa_2) \right. \pm \nonumber\\ 
\left.  vq_1(\kappa_1+\kappa_2)(m_0\pm E - b\kappa_1\kappa_2)\right]
\end{array}
\end{equation}    
where $\kappa_{1,2}=\kappa_{1,2}(0,E)$. Equation (\ref{Dirac_point_eq}) determines an energy dependence of the Dirac point on the values of the parameters $q_1$ and $q_2$. When the value of the diagonal parameter $q_1$ falls into two regions $q_{-}^{(1)} \leq q_1 \leq q_{+}^{(1)}$, $q_{-}^{(2)} \leq q_1 \leq q_{+}^{(2)}$, the Dirac point is removed from the bulk gap. In addition, the very SSs can vanish from the bulk gap at small momenta when the values of $q_1$ are those as illustrated in Fig.\ref{Fig:3Dspectrum}. We note that when the SSs absent in the bulk gap at small momenta, they cross the gap at edges of BZ in a such a way that the bulk-boundary correspondence to be valid (see Sec.\ref{lattice_model}). Meanwhile, $q_{\pm}^{(1,2)}$ have the following values: 
\begin{equation}
\begin{array}{l}
q_{-}^{(1,2)} = -\sqrt{\frac{m_0b}{2v^2}+\frac{1}{4}}\mp\sqrt{ \frac{m_0b}{2v^2}+\frac{1}{4} + \left( q_2+\frac{1}{2}\right)^2} \\
q_{+}^{(1,2)} = \sqrt{\frac{m_0b}{2v^2}+\frac{1}{4}}\mp\sqrt{ \frac{m_0b}{2v^2}+\frac{1}{4} + \left( q_2-\frac{1}{2}\right)^2}
\end{array}
\end{equation}
We stress that the values of $q_{\pm}^{(1,2)}$ are determined by not only bulk parameters $m_0,v$ and $b$ but also by the boundary parameter $q_2$. For other values of $q_1$, the Dirac point is in the bulk gap in the BZ center. 

If the boundary parameter $q_2$ is much larger than the other characteristic momenta (i.e. \mbox{$|q_2|\gg\max(1,(|m_0|b/v^2)^{1/2},|q_1|)$})  then the SS spectrum is quasi-linear with a small deviation of the Dirac point from the middle of the bulk gap 
\begin{equation}\label{large_q_spectrum}
E=\pm|v|k_{||}+\frac{2}{q_2}(m_0+bk_{||}^2).
\end{equation}  
The same spectrum (\ref{large_q_spectrum}) holds true when $q_2$ is replaced $q_1$ and when the parameter $q_1$ is the greatest in the BCs (\ref{BC_general}) \cite{bib:Menshov}. 

In the remainder of the section, we consider the Hamiltonian (\ref{3D_Ham}) and the general BCs (\ref{BC_general}) when the contribution of remote bands in (\ref{3D_Ham}) is small, i.e., $v^2\gg 4|bm_0|$. This case is simply a case of the standard 3D Dirac equation, for which a SS spectrum is described by the single phenomenological parameter $a_0$; see (\ref{DiracSSspectra}) and Ref.[\onlinecite{bib:VolkPin}]. Our aim is to express $a_0$ in terms of the three boundary parameters, $q_1,q_2$ and $q_3$. In this case, the wave function (\ref{3D_wave_function}) includes a smooth exponent with $\kappa_1(E)=-\sqrt{m_0^2-E^2+v^2k^2_{||}}/v$ and a rapid exponent with $\kappa_2=-v/b$. The former determines the spatial profile of the wave function and the latter only contributes to the BCs. Solving the dispersion equation with the above-mentioned values of $\kappa_{1,2}$ in leading order to $b$ (implying $|q_{1,2,3}|\gg b\kappa_1/v$) results in the SS spectrum (\ref{DiracSSspectra}), where
\begin{equation}\label{Dirac_limit}
a_0 = \frac{\left(2q_2+1 \right)^2+4q_3(q_1-1)}{\left(2q_2-1 \right)^2-4q_1(q_3-1)}.
\end{equation}
From equation (\ref{Dirac_limit}), we can conclude that in this limit ($v^2\gg 4|bm_0|$), the SSs exist in the bulk gap only at (i) $a_0>0$ if $m_0<0$ and (ii) $a_0<0$ if $m_0>0$ (see (\ref{DiracSSspectra})). However, in the case (ii) the massless SSs are not protected by the topological arguments as the Hamiltonian (\ref{3D_Ham}) describes a trivial insulator phase for $m_0>0,b>0$. 


\section{tight-binding model of 2D TI}\label{lattice_model}

Here, we study how general BCs effect ES spectra in a simplified tight-binding model of 2D TI in a similar way as it was done in Ref.[\onlinecite{bib:Isaev}]. This model allows us to calculate ES spectra over the entire BZ. We consider two-dimensional square lattice with lattice constant $a=1$ and four states per atom \cite{bib:Bern} $|s,\uparrow \rangle, (1/\sqrt{2})|p_x+ip_y,\uparrow \rangle,|s,\downarrow\rangle, (1/\sqrt{2})|p_x-ip_y,\downarrow\rangle$, where $\uparrow(\downarrow)$ means spin-up (spin-down) states. Neglecting terms that break e-h symmetry, one could obtain the Hamiltonian of the tight-binding model:
\begin{eqnarray}
	&& H = \Psi_i^{\dag}\left[\sigma_0\otimes\left((m_0-4b)\tau_z - 4d\tau_0 \right)\right]\Psi_i - \nonumber \\
	&& i \frac{v}{2}\left[\Psi_i^{\dag}\left(\sigma_0\otimes\tau_y e^{i\theta_i^y}\Psi_{x_i,y_i+1} +\sigma_z\otimes\tau_x e^{i\theta_i^x}\Psi_{x_i+1,y_i} \right) - \right. \nonumber \\
	&& \left. h.c. \right] + \sum\limits_\mu \left\{ \Psi_i^{\dag} \left[\sigma_0\otimes\left(b\tau_z + d\tau_0 \right)\right] e^{i\theta_i^\mu}\Psi_{i+\mu} + h.c.\right\},
\end{eqnarray}
where subindex $i = (x_i,y_i)$ enumerates sites of the square lattice, $i+\mu =\mu_i+1$ with $\mu = x,y$, $\theta_{i}^{\mu}$ is a U($1$) gauge field, $\Psi_i$ is four component operator, which annihilate electron on site $i$. It is well known that this tight-binding model describe a 2D topological insulator phase when $0<m_0/(2b)<4$, $m_0\neq 4b$ and a trivial insulator phase in the other cases.

We now consider a half-infinite square lattice that occupies $x_i>0$ with edge at $x_i=0$. Using $H$ we can derive expression for current in $x$ direction as variation of $H$ with respect to the gauge field $\theta_{i}^{x}$: 
\begin{eqnarray}
	&& j_{i}^{x} = \left. \frac{\delta H}{\delta \theta_{i}^{x}}\right|_{\theta_{i}^{x}=0} = i\left[\Psi_i^{\dag}\sigma_0\otimes\left(b\tau_z+d\tau_0\right) \Psi_{x_i+1,y_i} - h.c. \right] + \nonumber \\
	&& \frac{v}{2}\left( \Psi_i^{\dag} \sigma_z\otimes\tau_x \Psi_{x_i+1,y_i}  + h.c. \right)
\end{eqnarray}

 At the surface single-particle wave function $\psi_{x_i}$ obeys BC of the form
\begin{equation}\label{BC_appendix}
(ib\sigma_0\otimes\tau_z +\frac{v}{2}\sigma_z\otimes\tau_y)\psi_{1,0}=G\psi_{0,0},
\end{equation}
 where $G$ is a $4\times 4$ matrix, $\psi_{i}$ consists of four amplitudes of the corresponding orbitals. Vanishing $j^{x}$ at the surface fixes the form of the matrix $G$: $G=G^{\dag}$. Time-reversal symmetry and inversion result in 
\begin{equation}
G = \left(
\begin{array}{cccc}
g_1 & g_2 & 0 & -ig_3 \\
g_2 & g_4 & -ig_3 & 0 \\
0 & ig_3 & g_1 & -g_{2}\\
ig_3 & 0 & -g_2 & g_4
\end{array}
\right),
\end{equation}
where $g_{\overline{1,4}}$ are real parameters (which in general differ from the similar parameters in Sec. II). Further, we restrict ourselves by the case of block-diagonal matrix $G$ in spin space (i.e. $g_3=0$) and put $d=0$. This assumption will not influence main conclusions and allow us to study ES spectra separately for each spin.

\begin{figure}
	\includegraphics[width=8.5cm]{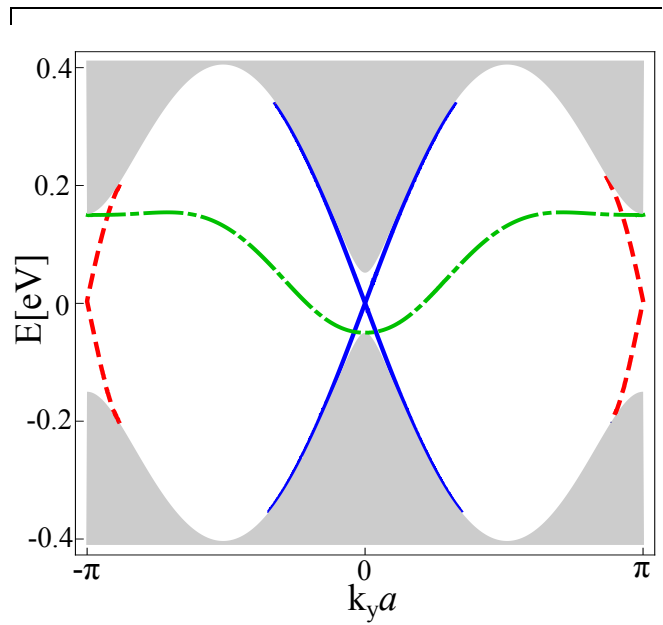}
	\caption{\label{Fig:SSspectrLattice2DTI}The electron spectrum $E(k_y)$ of semi-infinite 2D quadratic lattice. The shaded region corresponds to the continuous spectrum (bulk solutions). The edge states are described by solid curves for open boundary conditions ($g_1 = g_4 = \infty$, $g_2 = g_3 =0$), and dashed curves for $g_2 = 0.5$, $g_1 = 0.1$, $g_3=g_4=0$, and dash-dotted curve for $g_2=g_3=0$, $g4=\infty$, $g_1 = -0.05$. Tuning the boundary parameters modifies edge state spectra end even removes the Dirac point of ES (dash-dotted curve). However, boundary parameters cannot change the odd number of edge state pairs in the bulk gap in TI phase.}
\end{figure}

In order to study ES spectra we follow the method of analytical continuation of the bulk band structure \cite{bib:Davison}. The bulk spectrum of $H$ is
\begin{equation}
E=\pm\sqrt{\left (m_0 +2b(\cos k_x + \cos k_y - 2)\right )^2+v^2(\sin^2 k_x+\sin^2 k_y)}
\end{equation}  
The ESs possess complex $k_x=k+iq$ with $k$ and $q$ determined by condition $\Im E=0$. It leads to (i) $k=0$ and $k=\pi$, (ii) $\cos k \cosh q = 2bm_0 + 2b(\cos k_y - 2)/(v^2-4b^2)$. Only in the case (i) an energy lies inside the band gap. Therefore, a general wave function of spin-up ES is expressed by  
\begin{equation}\label{gen_wave_function_appendix}
\begin{array}{l}
\psi_{i} = C_1
\left (
\begin{array}{l}
-E-M_1 \\
iv(\sinh q_1 -\sin k_y)
\end{array}
\right )e^{-q_1x_i + ik_yy_i} + \\
\qquad \quad C_2\left (
\begin{array}{l}
-E-M_2 \\
-iv(\sinh q_2 +\sin k_y)
\end{array}
\right )e^{i\pi x_i-q_2x_i+ik_yy_i}
\end{array}
\end{equation}
where $M_{1,2} = m_0+2b(\pm\cosh q_{1,2}+\cos k_y -2)$, $C_{1,2}$ are arbitrary coefficients. Inverse decay lengths $q_{1,2}>0$ are determined by the energy and the momentum $k_y$:
\begin{equation}
\begin{array}{l}
\cosh q_{1,2} = \frac{\pm (2bm_0+4b^2(\cos k_y -2) )}{v^2-4b^2}+ \\
\\
\frac{ \sqrt{v^2(m_0+2b(\cos k_y -2))^2+(v^2-4b^2)(v^2(\sin^2 k_y+1)-E^2) } }{v^2-4b^2}
\end{array}
\end{equation}  
Dispersion equation results from satisfying the wave function (\ref{gen_wave_function_appendix}) the BC (\ref{BC_appendix}). For the spin-up wave function it yields
\begin{widetext}
\begin{equation}\label{lattice_dispersion}
\begin{array}{l}
\left[ \left( g_1-Be^{-q_1}\right) \left(E+M_1 \right)-2A(\sinh q_1 - \sin k_y) \left( Ae^{-q_1}+g_2\right) \right] \left[ \left ( g_2-Ae^{-q_2}\right) \left (E+M_2 \right )+2A(\sinh q_2 + \sin k_y) \left ( g_4-Be^{-q_2}\right )\right] + \\
\left[ \left ( Ae^{-q_1}-g_2\right ) \left (E+M_1 \right )+2A(\sinh q_1 - \sin k_y) \left ( Be^{-q_1}+g_4\right )\right] \left[ \left ( Be^{-q_2}+g_1\right ) \left ( E+M_2 \right )-2A(\sinh q_1 + \sin k_y) \left ( Ae^{-q_2}-g_2\right )\right] = 0,
\end{array}
\end{equation}
\end{widetext}
Dispersion equation for spin-down ESs is given by the same equation (\ref{lattice_dispersion}) with replacement $k_y\to -k_y$. In Fig.4 we plot spectra of the ESs in the topological insulator phase. Tuning the boundary parameters $g_1,g_2,g_4$ we can strongly effect the ES spectra (see Fig.\ref{Fig:SSspectrLattice2DTI}). It is even possible to destroy the massless ESs (dash-dotted curve) or remove it from the BZ center (dashed curves), however, the boundary parameters cannot change the hallmark of topological insulators - the odd number of ES pairs in the bulk gap. It is the main result of this section.

\section{Conclusions}

We derived the BCs for the envelope functions and found SS (ES) spectra of semi-infinite 2D TI based on an  HgTe/(Hg,Cd)Te quantum well and a 3D TI of Bi$_2$Se$_3$-type. In the 2D TI with the Hamiltonian (\ref{2DHam}) and the bulk parameters from \cite{bib:Qi}, the open BCs may be not suitable for the small Hamiltonian parameters $b$ and $d$. In that case, we showed that it is more appropriate to use the one-parameter BCs (\ref{DiracBC}) for small $k$. In the general case (when $b$ and $d$ are large enough), the BCs contain six real phenomenological parameters, two of which equal zero if a boundary does not mix the spins of an electron or if it possesses spatial inversion. In frames of the envelope function approximation we showed how boundary parameters modify spectra of SSs in the center of BZ. Some values of the parameters move the Dirac point in the SS spectra to the edges of BZ, other values destroy the massless SSs. Using the tight-binding model of 2D TI we revealed that boundary parameters cannot violate the bulk-boundary correspondence, i.e. there is always odd number of SS pairs in the bulk gap, see also \cite{bib:Persh}. But boundary parameters may strongly affect the SS spectra. In addition, we showed that non-protected massless SSs may be in the gap even in normal insulator phase ($m_0 > 0$).

At (111) surface of Bi$_2$Se$_3$, the BCs are described by the three real parameters. The SS spectrum again depends a great deal on the parameter values. When the parameters are large compared with the decay length of the SSs (e.g., in the case of the open BCs), the Dirac point of the SS spectrum is nearly in a middle of the gap in the BZ center with almost linear dispersion. We found regions of the boundary parameter values for which the SSs are pushed out the gap for small momenta. This means that they cross the bulk gap at large momenta for these boundary parameters to make bulk-boundary correspondence valid. 

We are grateful to Liang Fu for fruitful discussion. This work was supported by RFBR grants \#14-02-31592, and \#14-02-01166. The work of VVE was supported by the Dynasty Foundation.

\appendix

\end{document}